# StochSoCs: High performance biocomputing simulations for large scale Systems Biology

Elias S. Manolakos
Department of Informatics and Telecommunications
National and Kapodistrian University of Athens
Athens, Greece
eliasm@di.uoa.gr

Elias Kouskoumvekakis
Department of Informatics and Telecommunications
National and Kapodistrian University of Athens
Athens, Greece
eliask@di.uoa.gr

**INVITED PAPER**

*Abstract*— The stochastic simulation of large-scale biochemical reaction networks is of great importance for systems biology since it enables the study of inherently stochastic biological mechanisms at the whole cell scale. Stochastic Simulation Algorithms (SSA) allow us to simulate the dynamic behavior of complex kinetic models, but their high computational cost makes them very slow for many realistic size problems.  We present a pilot service, named *WebStoch*, developed in the context of our *StochSoCs* research project, allowing life scientists with no high-performance computing expertise to perform over the internet stochastic simulations of large-scale biological network models described in the SBML standard format. Biomodels submitted to the service are parsed automatically and then placed for parallel execution on distributed worker nodes. The workers are implemented using multi-core and many-core processors, or FPGA accelerators that can handle the simulation of thousands of stochastic repetitions of complex biomodels, with possibly thousands of reactions and interacting species. Using benchmark LCSE biomodels, whose workload can be scaled on demand, we demonstrate linear speedup and more than two orders of magnitude higher throughput than existing serial simulators.

## I. INTRODUCTION

Stochastic simulation of large-scale biochemical reaction networks has become an essential tool for Systems Biology. It enables the in-silico investigation of the dynamics of complex biological system under different conditions and intervention strategies, while also taking into account the inherent 'biological noise', that may play an essential role especially in the low species counts regime [1]. However, stochastic simulation presents a great computational challenge since in practice we need to execute thousands of repetitions of a complex simulation model to assess statistically the behavior of the underlying stochastic dynamical system it represents. The problem's computation work scales very rapidly in two dimensions, both with the number of repetitions *and* with the number of reactions in the model. The worst-case scenario when we need to execute tenths of thousands of repetitions of a  biomodel with thousands of reactions and species [2].

Our group has developed a novel web accessible stochastic simulation service, called *WebStoch* [3], which allows remote authorized users with no HPC expertise to perform efficient stochastic simulations of very large scale kinetic models supplied in standard SBML (Systems Biology Markup Language) [4] file format (biomodels) over the internet. The service parses the submitted biomodel and then executes user specified simulation jobs in the parallel processing back-end we have developed as part of the *StochSoCs* research project [5]. The users do not have to worry about how to configure the computations for optimal performance. In this way, simulation requests are very easy to setup and many of jobs can be submitted in an efficient and user-friendly manner.

The service is designed to support worker nodes implemented with heterogeneous hardware architectures such as many/multi-core x86 processors or Systems on Chip (SoC) accelerators that our group has designed as parametric IP cores and implemented for FPGA devices [5], [6]. The stochastic simulations that each worker executes can use either the First Reaction Method (FRM) SSA of Gillespie [7] or the Next Reaction Method (NRM) SSA of Gibson & Bruck [8]. Moreover, the service supports *Single Simulation in Parallel* (SSIP) execution style, in which a complex model's reactions are partitioned among the parallel workers, or *Multiple Simulations in Parallel* (MSIP), where each worker executes all model reactions but the stochastic repetitions of a simulation job are distributed evenly among the available workers. The mode of parallelism employed depends on user preferences and on the optimal configuration for the available hardware implementation of the worker node. Our service can handle efficiently simulation jobs with thousands of stochastic repetitions of a complex SBML biomodel with thousands of reactions and interacting species and is targeting large scale biocomputing for Systems Biology.

We have performed a multitude of experiments in order to assess the throughput and performance that each heterogeneous worker can deliver to the stochastic simulation of biomodels of increasing complexity. Using the developed by our group Linear Chain System Extended (LCSE) [5] benchmark biomodels, in which the workload and degree of reaction interdependencies can be increased in a controlled manner, we performed FRM and NRM simulations on parallel workers implemented using: a modern multi-core processor with 8 cores clocked at a frequency of 3 GHz (Intel Core-i7 5960X), a many-core processor with a Network on Chip (NoC) that contains 48 legacy Pentium cores clocked at 533 MHz (Intel Single-chip Cloud Computer – SCC [9]) and our group's power efficient stochastic simulation FPGA SoCs with up to 32 cores clocked at frequencies of up to 220 MHz. Experimental results demonstrate that all architectures considered have the potential to deliver high performance and linear speedup as the problem scales on both dimensions i.e. in terms of the biomodel's complexity (number of reactions) as well as in terms of the number of stochastic repetitions required for the simulation job.

The need to exchange biomodels between members of the scientific community as well as between different tools in a computational pipeline has created the need for uniform coding and a common standard for their description. Our system can accept, parse and then simulate biomodels expressed in the Systems Biology Markup Language (SBML) which has become the defacto standard in the community. The popularity of SBML is evidenced by the fast-growing number of models deposited to databases such as the European Bioinformatics Resource EMBL-EBI Biomodels DB [10]. SBML enables standardized coding of reactions, molecular species, reaction constants and parameters, as well as of the mathematical rules that govern species interactions, into machine readable (text) XML format files called "SBML biomodels".

One such biomodel was developed by our group and has been selected as "model of the month" (3/2015) of the EMBL-EBI Biomodel Database [11]. It can be used to study how the oligomerization of the protein Alpha-synuclein (ASYN) affects different parts of the cell and disturbs the homeostasis of dopaminergic neurons, a process that is believed to play a key role at the onset of Parkinson's disease [12]. Our ASYN biomodel has *m=136* reactions (mass action kinetics), *n=90* species and is a typical (medium size) biomodel of the EBI Biomodels DB, with a mix of reactions of different orders and stoichiometries.

The rest of the paper is organized as follows: In Section II we review the current landscape of HPC for stochastic simulation of reacting systems and present a summary of our *StochSoCs* research project achievements. In Section III we present some basic knowledge on simulations of biochemical reaction networks for the non-expert reader. In Section IV we discuss the performance and scalability evaluation of the heterogeneous architectures that our *WebStoch* service supports. In Section V we discuss the current status of *WebStoch* and point to future improvements. Finally, in Section VI, we close by summarizing our findings.

## II. HPC FOR STOCHASTIC SIMULATIONS

Modeling an organism's dynamic behavior at the cellular level may require thousands of reactions. A typical example is the metabolic network E-Coli [13] which contains 1,260 molecular species and 2,077 reactions. As reconstructions of whole species metabolic networks and tools for estimating their kinetic parameters are rapidly emerging [14], the need for the efficient stochastic simulation of large-scale biochemical reaction networks to study their stochastic dynamics will become more and more pressing.

The second dimension along which the complexity of stochastic simulation problem scales is the number of repetitions (realizations) required to approximate the stochastic behavior of a biological network as the number of its reactions increases. Modern software platforms developed to simulate biochemical reaction networks are throughput-limited since they fail to calculate more than half a million simulation steps per second, even if these are placed on latest generation processor units. From all the above, it is clear that the simulation of entire cells and cell populations (tissues) will remain an elusive goal unless we develop new computer systems that will cope with the fast-increasing computational needs of Systems Biology [15]. This will require that effective algorithms, specialized computer hardware accelerators, heterogeneous computing, are all employed in a concerted effort to address the stochastic simulation 'speed challenge'.

Power efficient hardware accelerators for stochastic simulation exploit the fine grain parallelism afforded by modern FPGA devices. However, the design of such aggressively pipelined Systems on Chip (SoC) is still a cumbersome process where the designer should strike a good balance between the complexity of the class of biomodels supported by the SoC and the size of the FPGA device used in terms of on-chip resources (LUTs, RAMs and DSPs). Examples of FPGA solutions are those in [16], [17] and [6]. Likewise, GPU based solutions, such as those in [18], [19] and [20] exploit the massively parallel compute power of GPUs that are nowadays readily available on the average scientist's desktop PC. Their major drawback lies however in the difficulty of generating in an efficient and automated manner 'kernel' (software) implementation for every SBML biomodel at hand, and in the amount of available on-chip fast access RAM memory (hardware), which is very limited for the needs of stochastic simulations of networks with a large number of reactions and species.

In addition, there exist pure software stochastic simulation solutions for today's commodity multi-core CPUs, but these certainly lack the performance and power efficiency of the above-mentioned hardware accelerators. Advances in modern compiler technologies, in conjunction with the already perfected instruction level parallelism (ILP) of multi-core processors, help mitigate the lack of specialized parallel hardware. Software simulators exploit all these to provide decent performance on stochastic simulations of medium size biomodels (with up to few hundreds of reactions) using SSA algorithms. Some well-known software tools that support stochastic simulation are: COPASI [21], StochKit [22] and MATLAB's System Biology toolbox. We should also mention the distributed version of COPASI that uses the HTCondor job scheduler and is thus named Condor-COPASI [23]. This tool needs extensive configuration and setup and aims to harness the power of a cluster for deterministic and stochastic simulations. At its core, Condor-COPASI still uses the same feature-rich COPASI software that is however inherently serial in nature.

The objective of our *StochSoCs* research project was the design and implementation of a high performance computational resource (service) that will be able to simulate efficiently very large-scale biochemical reaction networks with thousands of reactions and interacting species. To the best of our knowledge, there is currently no open computing resource that can harness the power of the different heterogeneous architectures (multi / many-core CPUs, FPGAs, GPUs) and that can accept as input user queries (in the form of SBML biomodels) parse them, and simulate the extracted network stochastically for a large number of repetitions on these architectures, and finally return results (e.g. time series of species counts and statistics) back to the user who is not required to code anything whatsoever. Thus, for large-scale networks simulation, users are currently limited to developing their own model-specific solutions for CPUs or GPUs (programmed in C/C++, CUDA, MATLAB etc.), a time-consuming task that only expert HPC programmers can handle. To improve this situation and meet our objectives, we have set the following specifications:

*Flexibility*: Worker nodes should be flexible and parametric entities in terms of the characteristics of the biomodels they can handle (number of reactions *(m)*, number of molecular species *(n)*) and in terms of the underlying hardware architecture that can be heterogeneous (CPU, FPGA, GPU etc.), having different core numbers (*C*), threads supported by each core (*T*) etc.

*Biomodel Complexity:* The simulation tasks placed on the service should be able to deal with complex biochemical networks with thousands of reactions. Reactions that describe the biological system will be governed by kinetic laws (mass action) of up to third order, while taking into account all possible stoichiometries.

*Performance:* Each worker supported by the service should be able to compute millions or even billions of stochastic simulation steps per second while efficiently managing the large number of stochastic repetitions performed.

*Scalability:* The service should support multiple stochastic simulations in parallel using different classes of worker nodes. Moreover, all service workers should have scalable implementations depending on the simulation's workload.

*Ease of use:* The service should be easy to interact with and navigate by non-expert users who should not be required to have any knowledge of the underlying heterogeneous architectures backend performing the parallel simulation jobs.

With the above specifications in mind, we have designed and developed a web-accessible, easy to setup and deploy computing platform, called *WebStoch* [3], which allows System Biology scientists and researchers with just an internet connection, but no programming skills, to submit SBML biomodels for efficient large-scale stochastic simulation. Simulation jobs are handled efficiently because the user query is parsed, analyzed and executed using the worker backend that the user selects. Especially for fully parametric FPGA workers, we generated several pre-synthesized SoC implementations for the task at hand. Currently, a medium size Kintex 7 Xilinx FPGA can implement SoCs with up to $C = 16$ cores and is able to handle biomodels with up to $m = 4096$ (4K) reactions, each reaction being up to $3^{rd}$ order with as many as 5 products and mass action kinetics. This is the first attempt to produce a platform with a flexible FPGA backend for efficient stochastic simulations of large-scale biomodels submitted as SBML files, without requiring any effort whatsoever from the end-user. Currently the pilot resource is fully functional and accessible in [3]. We encourage research groups that want to use it, to provide feedback and possibly collaborate with us on systems biology research that will benefit from its use, to contact us for improvements or requests on features they might have based on their needs.

### III. SIMULATION OF BIOCHEMICAL REACTION NETWORKS

There are two frameworks to follow when describing the dynamic behavior of a biochemical reaction networks: The first is based on ordinary differential equations (ODEs), while the second and more realistic tries to mimic the way nature works (discrete reaction events) and also account for the intrinsic and extrinsic stochasticity ('noise') of biological systems by employing Stochastic Simulation Algorithms (SSA).

A mathematical model is called deterministic if its behavior is exactly reproducible. Biomodels described by ODEs fall into this category and are assumed to represent the dynamic behavior of an "average" cell. Although they can be used to assess the transient and steady state average behavior of a biological system's state they cannot be used to study the effects of biological noise, an important limitation when modeling cellular mechanisms at the low species count regime.

Unlike deterministic modeling methods, stochastic processes aim to estimate the probability of finding the system in a particular state at a particular time. In addition, if the transition probability of the system depends only on its present state, then the system can be modeled with Markov chains. These are a special type of Markov processes that have either discrete state space or discrete index set (often representing time) allowing the system's stochastic behavior to be approached in a principled manner using probability theory. Employing Markov processes as the stochastic model for biochemical reaction networks allows us to approximate their time evolution as a sequence of discrete steps (reaction events).

Stochastic Simulation Algorithms (SSAs) employ random numbers generation to approximate the stochastic evolution of a dynamic system's state as a sequence of triggered reaction events that transition the underlying biological network from one discrete state to the next in a principled manner. Monte Carlo (MC) experiments are also used to repeatedly execute these algorithms utilizing different random number sequences each time, while also possibly changing their initial species conditions or reaction parameters at the start of each stochastic repetition, to obtain an ensemble of species vs. time trajectories. SSAs can describe with fidelity the npn-deterministic behavior of a true biological system taking into account its native thoughtfulness. The disadvantage of this approach lies however in the very nature of MC experiments that require stochastic execution for thousands of repetitions in order to statistically approximate the non-deterministic behavior of the underlying system [24].

Each fundamental reaction event can be described as the transition of the network from its present state *w* to a new state

$w'$. If $P(w,t)$ is the probability that the system is in state $w$ at time $t$, it can be shown that the evolution of probability $P(w,t)$ is governed by the following equation:

$$\frac{\partial P(w,t)}{\partial t} = \sum_{w' \in W,\ w' \neq w} T(w' \to w)\, P(w',t) - T(w \to w')\, P(w,t)$$

where the sum extends over the entire state space and can describe the dynamic evolution of any stochastic system. The transition function $T(w \to w')$ gives the probability density per unit time to switch the system from state $w$ to state $w'$. This is the well-known Chemical Master Equation (CME) [24]; its first term essentially expresses the probability of the system going from other states $w'$ back to the present state $w$, while the second term expresses the reverse, i.e. the probability that the system goes from the present state $w$ towards other states $w'$. Unfortunately, the CME does not have analytical or numerical solutions, except for a few cases of very simple systems, since it will have to be solved for each different molecular species. Moreover, the attempt to solve a set of CMEs is computationally inefficient and usually impossible given normal time constraints.

In general, a biochemical reactions network model is composed of $n$ species $\{S_1, ..., S_n\}$ that may interact through $m$ reaction channels $\{R_1, ..., R_m\}$. To simplify the analysis, we consider that all species are uniformly distributed within some volume $\Omega$. This assumption allows us to simplify the calculations by ignoring spatial inhomogeneity effects that exist in the real world. Let $X_i(t)$ be the concentration of species $S_i$ at time $t$. The state of the system at time $t$ is then $X(t) = (X_1(t), X_2(t), ..., X_n(t))$ with initial conditions $X_0(t) = x_0$ at initial simulation time $t = t_0$. The model can be described by a Markov chain, where the next state is only dependent on the present one. Simulating stochastically this model yields *trajectories* of the state $X(t)$. The most popular SSAs for biochemical reaction network models are the following: D.T Gillespie's original Direct Method (DM) [25] and his subsequent easy to parallelize First Reaction Method (FRM) [7], as well as Gibson and Bruck's Next Reaction Method (NRM) [8] optimized for best serial performance. The FRM and NRM are the two SSAs that our service currently supports.

An SSA executes as sequence of reaction cycles (RCs). In each RC of the FRM, all reactions ($m$) of the biomodel are visited and their propensity functions calculated based on the reaction kinetic laws. For mass action kinetics, a reaction propensity is proportional to the reactant species counts and a kinetic constant parameter modeling environmental effects (e.g. affinity conditions etc.). At the end of the RC, the SSA determines the "winner" reaction that will be executed next and its putative execution time using the calculated propensities and a randomizing strategy that destroys determinism so that reactions with small propensities can still have a chance to win. Using the winner reaction's ID and its putative time, the algorithm updates the system's state and advances the simulation time. The same stochastic race of reactions is repeated for as many RCs as needed for the system to reach the simulation time limit of the experiment ($T_{sim}$) set by the user.

The FRM SSA has the advantage that it can be fully parallelized by dividing its computation workload among the available processing units. Each RC workload can be processed in parallel by partitioning its $m$ reactions on the available $N$ processing units. This workload partitioning scheme will be referred to from now on as *Single Simulation in Parallel* (SSIP) and the associated workload placed on each processing unit equals to $W_{SSIP} = m / N$ reactions. Since we need to perform MC experiments with a lot of stochastic repetitions (realizations), say $R$, of a biomodel's simulation, we can instead distribute the $R$ independent repetitions among the $N$ processing units. This coarser grain workload partitioning scheme will be referred to as *Multiple Simulations in Parallel* (MSIP) and the corresponding workload for each processing unit will be $W_{MSIP} = R / N$ repetitions in this case. Unlike the FRM SSA, the NRM SSA can only be parallelized following the MSIP scheme. This is due to a priority queue shared data structure that the NRM maintains for storing what is called the *Dependencies Graph* (DG) of reactions and which prevents efficient parallelization. In [6] we have developed an aggressively pipelined NRM SoC for FPGAs that "hides" the access latency to the DG structure and achieves excellent scalability as the biomodel's DG edges increase.

IV. PERFORMANCE AND SCALABILITY EVALUATION

We will discuss next simulation experiments performed in order to assess the throughput and performance delivered, when our service is running the same biomodel but with different architectures and core configurations. We measured the running times and total number of reaction cycles (RCs) performed during all repetitions of each simulation job and report the simulation's Throughput (measured in *Mega Reaction Cycles per second - MRC/sec*), Performance (in *Mega Reactions per second - MR/sec*), and the Speedup ($S$) factor achieved relatively to our baseline(s), for a variety of core configurations. Note that during the execution of each reaction cycle when using the FRM the simulator should evaluate all $m$ reactions of the biomodel.

*A. Methods*

The Linear Chain System Extended (LCSE) biomodel with $2^{nd}$ and $3^{rd}$ order reactions [5] was used for all simulation experiments. This biomodel was developed by our group as a MATLAB script that generates standards compliant SBML biomodel files. It is suitable to use as a benchmark, in the sense that we can easily scale its workload on demand, both in terms of the number of reactions ($m$) and in terms of the average number of dependent reactions ($D_{aver}$). The latter feature is a unique property of this biomodel that is very useful in assessing the performance of the NRM SSA algorithm. The NRM is a better serial algorithm because in each RC it evaluates only the propensities that may change i.e. those of the reactions that depend on the winner reaction; this information is stored in the DG data structure. On the other hand, the NRM is much more difficult to parallelize [6].

The heterogeneous hardware architectures that our Simulation Worker employs for running stochastic simulations submitted through the service are the following:

*FPGA SoCs:* They have been developed as synthesizable parametric IP cores by our group and their most important

TABLE I. FRM EXPERIMENTS SETUP

| Architectures Compared: |  |
|---|---|
| 1. **StochSoCs SSA FPGA SoC**: 180 MHz, $C$ = 1-32 Cores (1 PE Each) <br> 2. **Intel Core-i7 5960X**: 3.0 - 3.5 GHz, $C$ = 1, 4 or 8 Cores <br> 3. **Intel SCC NoC**: 533 MHz, $C$ = 1, 12, 24, 36 or 48 Cores <br> **Speedup Baseline**: $C$ = 1 Core (for each of the above) | |
| **Parameters** | $T_{sim}$ = 1 sec, $T_{sam}$ = 0.1 sec, $R$ = 1 <br> $MIS$ (Max Internal Steps) = $10^9$ steps |
| **Biomodel Test Cases** | LCSE(m) Biomodel 2$^{nd}$ Order <br> $D_{aver}$ = 33 and $m$ = <br> - 512 (512 Reactions, 264 Species) <br> - 1K (1024 Reactions, 528 Species) <br> - 2K (2048 Reactions, 1056 Species) <br> - 4K (4096 Reactions, 2112 Species) |

features are presented in [5] (FRM SSA SoC) and [6] (NRM SSA SoC).

*An octa-core processor:* The Intel Core-i7 5960X (Extreme Edition) processor running at 3.0 GHz when workload is placed on all cores and up to 3.5 GHz, when workload is placed on a single core. It also contains 32GB of DDR4 RAM and a fast SSD with GNU/Linux O/S. We should note that the cores of this processor support the Intel Hyper-Threading (HT) technology in order to allow SMT execution with two threads on each core, however we decided not to use it since, besides complicating measurements (i.e. we cannot fully equate a quad-core processor as an "octa-thread"), it can sometimes limit the performance when the workload placed on all cores approaches 100%.

*The Intel Single-chip Cloud Computer (SCC):* This is an experimental many-core Network on Chip (NoC) processor architecture consisting of a 4 x 6 = 24 mesh of "tiles". Each tile includes two modified Pentium (P54C) cores, along with their 32KB L1 (divided equally for instruction and data) and 256KB L2 caches. It also contains a Mesh Interface Unit (MIU) with circuitry to connect the cores onto the network and allow them to run at different frequencies. In addition, a 16KB Message Passing Buffer (MPB) memory provides fast communication between the cores of the network, for a total MPB size of 384KB [9].

The last two processor based architectures initialize and execute the software framework we have developed for performing stochastic simulations on multi-core and many-core processor architectures, called *Hybrid Stochastic Simulator* (HSS). Our framework provisions the whole simulation flow, from configuration and loading of resources to the actual simulation scheduling and parallel execution, on the set of cores or threads the user has chosen when submitting the simulation job on the service. The supported SSA algorithms are currently the FRM and NRM in SSIP (only for FRM) and MSIP modes of parallel operation. The software framework can run on any modern workstation with a x86 processor and its original version was presented in [26].

TABLE II. FRM RESULTS (LCSE 2$^{ND}$ ORDER, M = 512 - 4096)

| Architecture | Throughput (MRC/s) | Performance (MR/s) | Speedup (vs C=1 Core) |
|---|---|---|---|
| StochSoCs FPGA SoC ($C$ = 32) | 1.28 – 0.79 | 653 – 3,229 | 4.50 – 18.84 |
| Intel Core-i7 5960X ($C$ = 8) | 0.053 - 0.007 | 27.20 – 28.54 | 5.62 – 5.82 |
| Intel SCC NoC ($C$ = 48) | 0.003 – 0.017 | 8.92 – 13.65 | 22.15 – 45.96 |

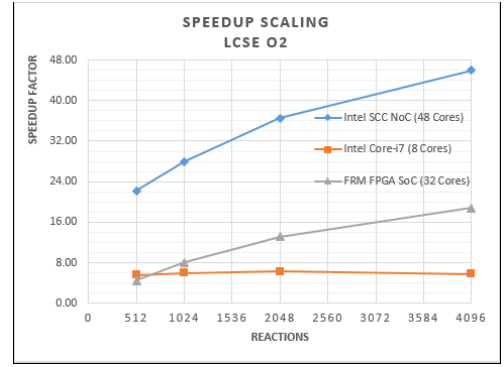

**Figure 1.** Speedup scaling for FRM SSA in SSIP mode. Intel SCC NoC (C=48), Intel Core-i7 5960X (C=8) and FRM FPGA SoC (C=32) as the number of reactions increases ($m = 2^{9..12}$) on LCSE 2$^{nd}$ order biomodels.

*B. FRM Results*

All simulation experiments in this section were performed using the FRM SSA in the SSIP (*Single Simulation in Parallel*) mode of operation. The baselines for benchmarking were established by running each simulation setup on a single core of the three different architectures presented in the previous sub-section. These baselines correspond to a serial execution of the FRM SSA of a given size. All SSIP experiments were setup for a single repetition of the network's simulation. Table I provides a summary of the architectures compared, the parameters of the experiments and the test cases performed. We used the LCSE benchmark biomodel in four different configurations of increasing reactions workload ($m$): 512, 1024 (1K), 2048 (2K) and 4096 (4K), while maintaining a fixed number of cores (different for each architecture) in the evaluation. These experiments aim to evaluate the communication cost impact in SSIP mode of parallel execution which partitions the reactions workload of the FRM SSA among the many cores of each architecture. This kind of simulation can tell us how scalable each architecture is and how its efficiency behaves as the problem size increases, i.e. how large of a problem size each architecture instance can handle efficiently before its speedup levels off.

Figure 1 shows how speedup factor behaves for the three different architectures as the complexity (number of reactions ($m$)) of the problem (stochastic simulation of the LCSE 2$^{nd}$ order biomodel) increases. The Intel SCC NoC achieves a near optimal speedup scaling as $m$ increases in the range $m$ = 512 up to $m$ = 4096. We observe that for very large $m$ = 4096 the NoC architecture ($C$=48) achieves an almost perfect speedup factor (~46) i.e. it is capable to reach efficiency close to unity as $m$ increases. A similar trend is evidenced for the FRM FPGA SoC architecture with efficiency exceeding 0.58 for $m$ = 4096. Interestingly, the Intel Core-i7, exhibits a pretty much constant speedup of ~6X across the range of problem sizes considered.

Our FRM SoC worker is the only application specific hardware worker for the FRM algorithm's acceleration and achieves very high throughput and performance. Its throughput when all 32 cores are utilized for $2^{nd}$ order LCSE biomodels lies between 0.8 and 1.2 MRC/s depending on the number of reactions ($m$). An equally high throughput of about 1.7 MRC/s is achieved for $3^{rd}$ order LCSE biomodels ($m$ = 512). These throughput figures translate to a very high performance of up to 3,200 MR/s and up to 1,500 MR/s for large $2^{nd}$ ($m$ = 4K) and $3^{rd}$ ($m$ = 1K) order biomodels respectively. Although the performance of this worker is much greater than all others, the speedup gains of the FRM FPGA SoC architecture start to decrease beyond the 16-core barrier. This is attributed to the fact that the FRM SoC contains a binary reduction (comparison) tree connecting all cores and used to find the winner reaction with the minimum activation time. This increases its clock cycle by an $O(logC)$ amount, where $C$ is the number of cores of the FRM FPGA SoC bitstream.

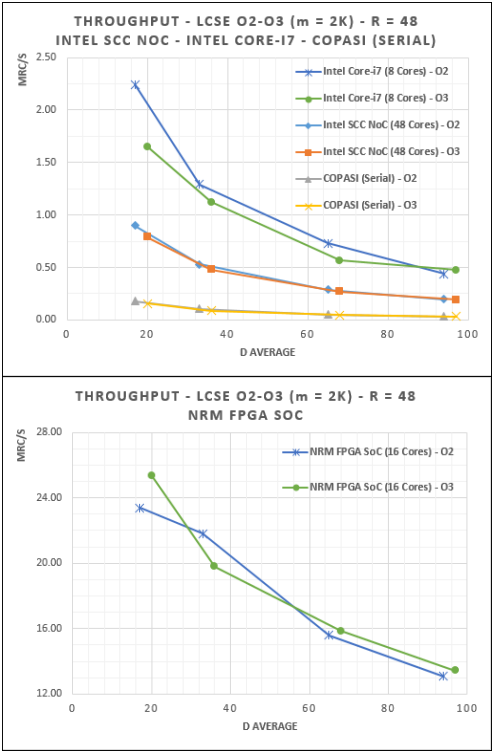

**Figure 2.** NRM SSA throughput comparison of the Intel SCC NoC ($C$=48), Intel Core-i7 ($C$=8) and NRM FPGA SoC ($C$=16 cores) with the serial simulator COPASI. LCSE 2nd and 3rd order biomodels were used with fixed $m$ = 2K, $R$ = 48 (MSIP) but increasing $D_{aver}$.

In summary (see also Table II), although the Intel SCC NoC and Intel Core-i7 architectures are inferior to the FRM SoCs, both in terms of throughput (MRC/s) and performance (MR/s) achieved for a given problem size, they have cores that communicate efficiently and thus their performance scales very well across the range of problem sizes considered. As these results suggest, especially many-core NOC based architectures present great promise for efficient stochastic simulations of very large biomodels using the parallelized version of the FRM SSA algorithm and software implemented workers.

TABLE III. NRM EXPERIMENTS SETUP

| Architectures Compared: | |
|---|---|
| **1. StochSoCs SSA FPGA SoC**: 160-190 MHz, $C$ = 16 Cores (1 PE Each) | |
| **2. Intel Core-i7 5960X**: 3.0 GHz, $C$ = 8 Cores (No Turbo) | |
| **3. Intel SCC NoC**: 533 MHz, $C$ = 48 Cores | |
| **Speedup Baseline**: Intel Core-i7 5960X: 3.5 GHz, $C$ = 1 Core (COPASI) | |
| **Parameters** | $T_{sim}$ = 1 sec, $T_{sam}$ = 0.1 sec $R$ = 48  <br> MIS (Max Internal Steps) = $10^9$ steps |
| **Biomodel Test Cases** | LCSE($m$) Biomodel $2^{nd}$ and $3^{rd}$ Order where $m$ = 2K and $D_{aver}$ = <br> - 17, 33, 65, 94 ($2^{nd}$ Order) <br> - 20, 36, 68, 97 ($3^{rd}$ Order) |

## C. NRM Results

All simulation experiments discussed here were performed using the NRM SSA algorithm in the MSIP (*Multiple Simulation in Parallel*) mode of operation. The baselines for benchmarking were established by running first each experiment using the serial simulator COPASI (single repetition). We then proceeded by running multiple repetitions ($R$=48) of the same experiment on each parallel. Table III summarizes the architectures and parameters used in the experiments. We utilized the LCSE benchmark biomodel, in both $2^{nd}$ and $3^{rd}$ order forms, keeping constant the number of reactions at a high value ($m$ = 2K) and increasing gradually the average number of reactions depending on the winner reaction ($D_{aver}$). Each core of a parallel architecture assumed an equal workload of $R / C$ repetitions. This kind of simulation was useful to assess the maximum performance an architecture can deliver, since after simulation initialization (where broadcasting of biomodel resource data occurs) the simulation repetitions proceed independently and there is no inter-core communication overhead involved.

The top plot of Figure 2 presents the throughput comparison of the three different architectures and COPASI (serial baseline) as the critical parameter $D_{aver}$ increases. We observe that the throughput of the Intel SCC NoC (48 Cores) and Intel Core-i7 (8 Cores) decreases as the complexity of the simulated biomodel increases (greater $D_{aver}$). The Core-i7 exhibits a maximum throughput of 2.24 MRC/s for LCSE $2^{nd}$ order with small $D_{aver}$ = 17 and 1.65 MRC/s for LCSE $3^{rd}$ order with $D_{aver}$ = 20. These figures drop to 0.43 MRC/s and 0.47 MRC/s for the larger $D_{aver}$ of 94 and 97 considered for $2^{nd}$ and $3^{rd}$ order biomodels respectively. A very similar behavior is observed for the Intel SCC NoC with a maximum throughput of about 0.9 MRC/s and a minimum throughput of 0.2 MRC/s for both LCSE $2^{nd}$ ($D_{aver}$ = 94) and $3^{rd}$ ($D_{aver}$ = 97) order experiments. The serial COPASI simulator delivers a much lower throughput (0.16 to 0.003

TABLE IV. NRM RESULTS (LCSE $3^{RD}$ ORDER, M=2K, $D_{AVER}$=20 – 97)

| Architecture | Throughput (MRC/s) | Performance (MR/s) | Speedup vs COPASI |
|---|---|---|---|
| StochSoCs FPGA SoC ($C$ = 16) | 25.38 - 13.43 | 508 - 1,302 | 161 - 428 |
| Intel Core-i7 5960X ($C$ = 8) | 1.65 - 0.47 | 33.02 - 45.98 | 10.47 - 15.10 |
| Intel SCC NoC ($C$ = 48) | 0.79 - 0.19 | 15.80 - 18.72 | 5.01 - 6.15 |
| COPASI Serial $C$ = 1 | 0.16 - 0.03 | 3.16 - 3.01 | 1 |

MRC/s) when run on a single Intel Core-i7 core. The results for the more demanding 3rd order LCSE biomodels are summarized in Table IV.

The bottom plot of Figure 2 demonstrates the very high throughput delivered by the NRM FPGA SoC for LCSE 2$^{nd}$ and 3$^{rd}$ order biomodel simulation experiments. It is evident that this architecture also loses performance as $D_{aver}$ increases for fixed number of reactions. The highest throughput of 23.37 MRC/s is observed when running the simplest LCSE 2$^{nd}$ order biomodel ($D_{aver}$ = 17). For the LCSE 3$^{rd}$ order biomodel with $D_{aver}$ = 20 we reached a throughput of 25.38 MRC/s which is the highest of all the NRM experiments regardless of architecture. On the opposite side, the lowest throughput is 13.07 and 13.43 MRC/s for LCSE 2$^{nd}$ order ($D_{aver}$ = 94) and LCSE 3$^{rd}$ order ($D_{aver}$ = 97) respectively. This kind of throughput is almost an order of magnitude (10X) greater than the octa-core Intel Core-i7 5960X something not surprising since we are using application specific accelerator hardware.

Let us now discuss the speedup achieved by the three architectures relatively to the corresponding serial COPASI software simulation. The top plot of Figure 3 presents the speedup of the Intel Core-i7 (8 Cores) and the Intel SCC NoC (48 Cores) vs. COPASI for both 2$^{nd}$ and 3$^{rd}$ order LCSE biomodels. The Intel Core-i7 achieved more than an order of magnitude (10X-15X) higher throughput than COPASI. Less speedup (5X-6X) was achieved by the Intel SCC NoC using less potent Pentium 4 processors. While our software based simulators running on the Intel SCC NoC and the Intel Core-i7 demonstrated substantial speedup gains vs. the serial COPASI simulator, the NRM FPGA SoC completely outperformed all of them, delivering throughputs of two orders of magnitude higher relatively to COPASI. Specifically, as shown in Figure 3 (bottom) it could deliver up to 367X and up to 428X higher throughput for 2$^{nd}$ and 3$^{rd}$ order LCSE models respectively. This demonstrates the very high efficiency of our parametric SoC implementations and their optimal scaling characteristics when the repetitions workload is split evenly among the cores of the SoC. Using readily available larger FPGA devices could raise this speedup factors even more by using SoCs with more cores (NRM workers) operating in MSIP mode.

V. WEBSTOCH CURRENT STATUS & FUTURE WORK

The *WebStoch* service at its current version allows remote authorized users to execute flexibly stochastic simulations of large scale biochemical kinetic networks on heterogeneous architectures such as multi / many-core CPUs and FPGAs. The improvements that can be made in such a system that combine both hardware and software components are many. Our current efforts are focused on adding more types of worker nodes. We already have an 'alpha' prototype of a GPU worker that uses specialized kernels for stochastic simulation and has already shown promising scalability results. Multiple FRM repetitions can be executed in parallel and each repetition's workload also parallelized using the plethora of available GPU cores implementing a kernel. We are also working on a new Intel Xeon Phi worker development to exploit the many-core powerful architectures of Intel SCC's successors, codenamed 'Knights Landing' and 'Knights Corner'.

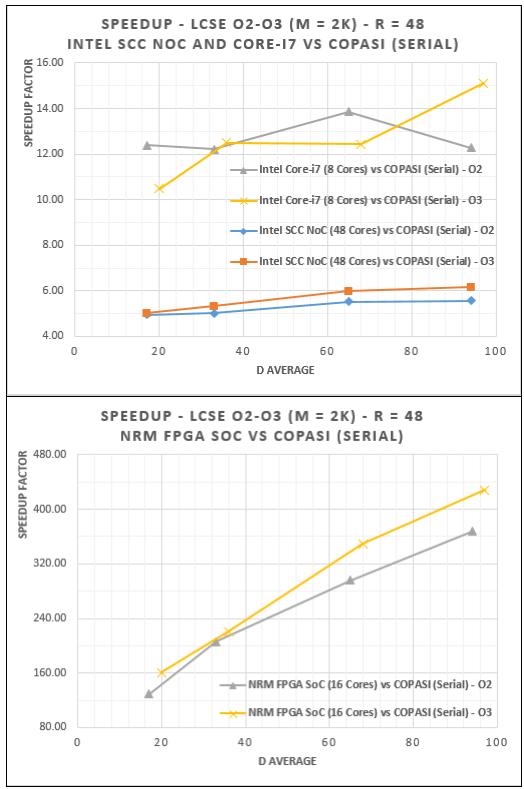

**Figure 3.** NRM speedup of Intel SCC NoC (48 Cores), Intel Core-i7 (8 Cores) and NRM FPGA SoC (16 Cores) vs. the serial simulator COPASI as $D_{aver}$ increases. LCSE 2$^{nd}$ and 3$^{rd}$ order models, $m$ = 2K, $R$ = 48.

Moreover, each of the existing workers and service components could be improved to give the ability to perform stochastic simulations with more features available to the users. Finally, we are also examining the possibility of deploying the full service to a compute cloud in order to exploit the massive scalability of cloud solutions that as of today support most of our current and future heterogeneous hardware architectures (CPU, FPGA, GPU).

VI. CONCLUSIONS

As part of the *StochSoCs* research project [5] we have designed, developed and deployed the a prototypical HPC service for parallel stochastic simulations of large scale biological networks that is accessible over the Internet. The service allows registered users to submit SBML biomodels of biochemical reaction kinetic networks along with the desired parameters for their stochastic simulation. The submitted SBML biomodel files are parsed automatically and then stochastic parallel simulations are initiated over the available workers in either SSIP or MSIP mode of parallel processing. Upon simulation completion, the server returns all species trajectories and simulation job statistics to the user for further analysis. Such a unique service provides a very useful resource that was not available before to the Systems Biology scientific community. It enables the efficient scalable stochastic simulation of large complexity biomodels without the user having to own or code fast processors or design dedicated hardware. The service has been successfully tested with a variety of SBML biomodels.

To evaluate the performance of the different heterogeneous workers we constructed synthetic (benchmark) biomodels in which the number of reactions ($m$) and/or the number of affected reactions by the winner ($D_{aver}$) can be increased in a controlled manner. Through a multitude of simulation experiments we have shown that both the Intel SCC NoC and Intel Core-i7 5960X processors exhibit very good scaling characteristics, as their x86 ISA cores can efficiently meet the increasing computational demands of simulation as the complexity of biomodels increases. In the case of the FRM SSA algorithm, their efficiency appeared nearly optimal and higher than that of the FRM FPGA SoCs, though of course their throughput (MRC/sec) and performance (MR/sec) was much smaller.

Additionally, it was confirmed that the performance of the parametric SoCs we have designed for the NRM SSA is practically independent of the biomodel's size ($m$), thus allowing the efficient simulation of very large biomodels. Comparing the NRM FPGA SoC (16 Cores) to the popular COPASI software simulator (serial), we demonstrated that as the number of affected reactions increases (and thus the biological network is getting more complex) the former can achieve up to 367 and 428 times better throughput than the latter when simulating LCSE benchmark models with 2$^{nd}$ and 3$^{rd}$ order reactions respectively. These results, combined with those discussed above for the FRM algorithm, clearly confirm the superiority of application specific accelerators implemented using FPGAs, although admittedly their design is much harder than software development targeting x86 architecture processors.

Overall, we believe that our research has fully met its objectives, contributing heterogeneous hardware and software system implementations for the efficient stochastic simulation of biological reaction networks of increasing complexity. This is indispensable for reaching the much bigger goal of reliable and realistic simulation of whole cell models, which is one of the main computational challenges of our decade and perhaps the "holy grail" of computational Systems Biology.

ACKNOWLEDGMENTS

This research has been supported in part by an "ARISTEIA II" project (PI: E.S. Manolakos) co-funded by the EU (European Social Fund) and GSRT national resources.